\documentclass[aps,prl,twocolumn,altaffilletter,nolongbibliography,numerical,flushbottom,secnumarabic,superscriptaddress,floatfix,nobalancelastpage]{revtex4-2}

\usepackage{graphicx}
\usepackage{braket}
\usepackage{booktabs}
\usepackage{array}
\usepackage{amsmath, amssymb}

\usepackage{bm}
\renewcommand{\vec}{\bm}

\usepackage[breaklinks, pdftex, hyperfootnotes=true, pdfpagelabels, bookmarks, pageanchor]{hyperref}
\hypersetup{colorlinks=true, pdfstartpage=1, pdfstartview=FitW, pdfborder={0 0 0}, breaklinks=true, pdfpagemode=UseNone, pageanchor=true, pdfpagemode=UseOutlines, plainpages=false, bookmarksnumbered, bookmarksopen=true, bookmarksopenlevel=1,hypertexnames=true, pdfhighlight=/O, urlcolor=blue, linkcolor=blue, citecolor=blue}

\usepackage{orcidlink}

\begin{document}

\title{Nonstabilizerness of Permutationally Invariant Systems}
\author{G.~Passarelli\,\orcidlink{0000-0002-3292-0034}}
\email{gianluca.passarelli@unina.it}
\affiliation{Dipartimento di Fisica, Universit\`a di Napoli ``Federico II'', I-80126 Napoli, Italy}

\author{R.~Fazio\,\orcidlink{0000-0002-7793-179X}}
\affiliation{The Abdus Salam International Center for Theoretical Physics,  34151 Trieste, Italy}
\affiliation{Dipartimento di Fisica, Universit\`a di Napoli ``Federico II'', I-80126 Napoli, Italy}

\author{P.~Lucignano\,\orcidlink{0000-0003-2784-8485}}
\affiliation{Dipartimento di Fisica, Universit\`a di Napoli ``Federico II'', I-80126 Napoli, Italy}

\begin{abstract}
Typical measures of nonstabilizerness of a system of $N$ qubits require computing $4^N$ expectation values, one for each Pauli string in the Pauli group, over a state of dimension $2^N$. For permutationally invariant systems, this exponential overhead can be reduced to just $O(N^3)$ expectation values on a state with a dimension $O(N)$. We exploit this simplification to study the nonstabilizerness phase transitions of systems with hundreds of qubits.
\end{abstract}

\maketitle

\textit{Introduction} --- Entanglement, by itself, is not sufficient to achieve quantum advantage. States with low entanglement can be represented with polynomial effort using tensor networks~\cite{vidal2004,verstraete2007,schollwock2011}, but there also exist highly entangled states that can be efficiently manipulated using classical computers~\cite{gottesman1998,gottesman1998-2}. On a quantum computer, any $N$-qubit evolution can be arbitrarily approximated using a combination of four types of unitary operators, the two-qubit CNOT gate and the one-qubit Hadamard $H$, $\pi/2$ phase shift $S$, and $\pi/4$ phase shift $T$ gates~\cite{nielsen2010}. The subset $\lbrace H, S, \mathrm{CNOT} \rbrace$ is nonuniversal, and generates the Clifford group $\mathcal{C}_N$. States that can be built by only applying gates in $\mathcal{C}_N$ on the computational basis states are known as stabilizers and they admit an efficient classical representation regardless of their entanglement~\cite{gottesman1997,gottesman1998,aaronson2004}.

Adding the $T$ gate to the ones in $\mathcal{C}_N$ allows for universal quantum computation. States that are built using unitaries from the $\text{Clifford}+T$ group have an extra resource known as nonstabilizerness, or, colloquially, ``quantum magic''~\cite{bravyi2005,howard2014,veitch2014,chitambar2019,seddon2019,zhou2020}, which renders them classically intractable when their entanglement is beyond the capabilities of tensor networks. This nontrivial interplay between entanglement and quantum magic is believed to be the key to unlock quantum advantage. For this reason, quantifying the amount of nonstabilizerness has become a central pursuit, not only in quantum information theory, but also in related areas such as condensed matter and statistical physics~\cite{liu2022}.

While there exist many universally accepted entanglement measures~\cite{nielsen2010,horodecki2009}, measures of nonstabilizerness are still under scrutiny. In general, the nonstabilizerness of a state can be measured as its distance from the set of stabilizers~\cite{veitch2014,howard2014-2,bravyi2016,bravyi2016-2,bravyi2019,heinrich2019,wang2019,wang2020,heimendahl2021,jiang2023,haug2023-3}, but quantities based on this definition are often difficult to compute. More appealing options are based on the analysis of the properties of the expectation values of all Pauli strings over the quantum state, the so-called Pauli spectrum. In fact, the Pauli spectrum of stabilizer states has precise properties~\cite{aaronson2004}, and any deviation is a symptom of nonstabilizerness~\cite{turkeshi2023-1}. Among these measures, we cite the stabilizer nullity~\cite{beverland2020}, the average entanglement-spectrum flatness over Clifford orbits~\cite{turkeshi2023,tirrito2023}, and the stabilizer $k$-R\'enyi entropies (SREs)~\cite{leone2023}.

Evaluating the above quantities, however, is extremely challenging, and the investigation of their properties so far has been limited to systems of $N \sim 10 $ qubits~\cite{niroula2023,leone2023} even in the case of integrable systems~\cite{oliviero2022,rattacaso2023}, with only few exceptions~\cite{haug2023-2,lami2023,fux2023,tarabunga2024}. In fact, the dimension of the Hilbert space of a system of $N$ spins grows exponentially as $D = 2^N$, and the number of Pauli strings $\hat P = \hat \sigma_1^{\alpha_1} \otimes \dots \otimes \hat\sigma_N^{\alpha_N} $, with $\alpha_j \in \lbrace 0, x, y, z \rbrace$ in the Pauli group $\mathcal{P}_N$ grows as $D^2 = 4^N$. This exponential complexity in both the state dimension and the number of expectation values makes it impossible to compute the nonstabilizerness of large systems. This is especially relevant since many-body systems have been shown to host a plethora of exotic entanglement-related phenomena~\cite{amico2008,laflorencie2016,lunt2022,passarelli2023} and recent works started investigating these types of phenomena also in magic~\cite{leone2023,niroula2023,fux2023,bejan2023,tarabunga2023-3}. It is therefore essential to find systems where the nonstabilizerness can be computed efficiently for large $N$. 
An important step forward in this direction has been achieved when the quantum state can be efficiently represented by tensor networks~\cite{haug2023-2,lami2023,fux2023,tarabunga2024}. It would be also desirable to find non-trivial many-body systems that admit an exact solution. In this work we focus on systems that are symmetric under the permutation group~\cite{chase2008,shammah2018}.
In this case one can exploit the symmetry to reduce the interesting portion of the Hilbert space to a more manageable subspace, allowing the study of much larger systems than normally accessible. For permutationally invariant systems, it is possible to compute the nonstabilizerness measures based on the Pauli spectrum with only a polynomial effort rather than an exponential one~\cite{moroder2012,novo2013}. This allows studying the nonstabilizerness phase diagram for this broad class of systems for large $N$.
After introducing some nonstabilizerness measures, we will discuss how permutational invariance allows to simplify their evaluation and, as an example, apply these findings to the Lipkin-Meshkov-Glick (LMG) model~\cite{lipkin1965}.

\textit{Definitions} --- The stabilizer R\'enyi entropies are extensively studied as probes of quantum magic. Entropies with R\'enyi index $ k < 2 $ are non-monotone under measurements in the computational basis followed by conditioned Clifford operations~\cite{haug2023}, which is why we will focus on the 2-R\'enyi entropy, the simplest one without this issue~\footnote{The investigation on this matter is still ongoing, see Ref.~\cite{haug2023}.}, and refer to it as SRE. For pure states, the SRE is defined as 
\begin{equation}\label{eq:sre}
    \mathcal{M}_2(\ket{\psi}) = -\ln \biggl[{\sum_{\hat P\in\mathcal{P}_N} \Bigl(\braket{\psi | \hat P | \psi}^2/D\Bigr)}^2\biggr] - \ln D.
\end{equation}
We recall that Clifford operators map each string of Paulis to a single string of Paulis. 
Given a pure-state density matrix $\hat \rho = \ket{\psi}\bra{\psi}$, the state $\ket{\psi}$ is a stabilizer if and only if $\hat\rho$ has $2^N$ nonzero components over the strings of the Pauli group while the rest are zero~\cite{aaronson2004}. Therefore, the SRE measures the entropy of the probability distribution $\Pi_P = \mathrm{Tr}(\hat P\hat \rho)^2 / D$, shifted by that of a stabilizer state. As shown in Ref.~\cite{leone2022}, it is a good nonstabilizerness measure, and satisfies the bound $0 \le \mathcal{M}_2 < N \ln 2$, with $\mathcal{M}_2 = 0$ if and only if $\ket{\psi}$ is a stabilizer.

%

\textit{Permutational invariance} --- A system is permutationally invariant when is made up of $N$ identical qubits and its Hamiltonian is only expressed in terms of collective spin operators, $\hat S_\alpha^{(N)} = \sum_{j = 1}^{N} \hat \sigma^\alpha_j/2$.   
Its Hamiltonian remains the same after any relabelling of the particle indices. In this setting, the only states that matter are those that respect the permutation symmetry of the model and are built as a symmetric superposition of states with $n$ excitations, with $n = 0, 1, \dots, N$. These states form a basis of the $(N + 1)$-dimensional symmetric subspace and are known as Dicke states~\cite{shammah2018}. They read
\begin{equation}\label{eq:dicke}
    \ket{N, n} = {\binom{N}{n}}^{-1/2} \hat{\mathcal{S}}\left[\ket{0}^{\otimes (N - n)} \ket{1}^{\otimes n}\right],
\end{equation}
where $\hat{\mathcal{S}}$ is the symmetrizer operator. Many important states in quantum information can be expressed in the Dicke basis. For example, the GHZ state can be written as $\ket{\text{GHZ}} = (\ket{N, 0} + \ket{N, N}) / \sqrt{2} \equiv (\ket{00\dots 0} + \ket{11\dots 1}) / \sqrt{2}$, whereas the state $\ket{N, 1}$, with one excitation, corresponds to the $\ket{W} = (\ket{00\dots 01} + \ket{00\dots 10} + \dots + \ket{10\dots00}) / \sqrt{N} $ state. Moreover, permutational invariant systems have applications in quantum 
metrology and quantum error correction~\cite{ouyang2014,ouyang2016,ouyang2021,ouyang2022-2}.

Because of permutation symmetry, the expectation value of a Pauli string on a symmetric state cannot depend on the order of the Pauli operators appearing in the string, but only on the number of $X$, $Y$, $Z$ and $I$ gates in the string~\cite{ouyang2022}. Let us call these numbers $N_x$, $N_y$, $N_z$ and $N_0$, and group them together in the quadruple $\vec{N} = \lbrace N_x, N_y, N_z, N_0\rbrace$. Of course, one must have $N_x + N_y + N_z + N_0 = N$. Any summation over the elements of the Pauli group can then be split as follows for any function $f(\hat P)$, 
\begin{equation}\label{eq:pi-pauli-sum}
    \sum_{\hat P\in\mathcal{P}_N} f(\hat P) = \sum_{\vec{N}} g(\vec{N}) f\bigl(\hat P(\vec{N})\bigr),
\end{equation}
where the operator $\hat P(\vec{N})$ is the permutationally-symmetric representative and $g(\vec{N})$ is its degeneracy. The number of distinct representatives is given by~\footnote{There are $N$ labels $\alpha_j$ to be separated in $k = 4$ sets by putting $k - 1 = 3$ separators among them. There are $(N + k - 1)! = (N + 3)!$ ways of doing that, and permutations of identical labels ($N!$) and separators [$(k - 1)! = 3!$] have to be excluded, hence Eq.~\eqref{eq:pauli-dimension-symmetric}.} 
\begin{equation}\label{eq:pauli-dimension-symmetric}
    \mathcal{D} = \binom{N + 3}{3} = \frac{(N + 1) (N + 2) (N + 3)}{6}
\end{equation}
while their degeneracy is given by the multinomial coefficient
\begin{equation}\label{eq:deg}
    g(\vec{N}) = \binom{N}{N_x, N_y, N_z, N_0} = \frac{N!}{N_x! N_y! N_z! N_0!}.
\end{equation}

It is possible to explicitly construct the matrix representation of the symmetric representative given a quadruple $\vec{N}$, in the Dicke basis. The technical details are discussed in the Supplementary Material~\footnote{See Supplementary Material, where we provide technical details on the symmetric Pauli group, the study of a different symmetric model, other measures of nonstabilizerness simplified by permutation symmetry, and apply mean-field theory to the ground states of the analyzed models. It includes Refs.~\cite{sakurai,derrida1981,gross1984,bapst2012,filippone2011}.}. Here, we just report the final result,
\begin{align}\label{eq:mel-pauli}
    &\braket{N, m | \hat P(\vec{N}) | N, n} = {(-i)}^{N_y} \\[1ex] &\quad \times \sum_{n_x, n_y, n_z} {(-1)}^{n_y + n_z} \frac{\binom{N_x}{n_x} \binom{N_y}{n_y} \binom{N_z}{n_z} \binom{N_0}{n - n_x - n_y - n_z}}{\sqrt{\binom{N}{n}\binom{N}{m}}}\notag \\[1ex]
    &\qquad\times\delta_{m, n + (N_x - 2 n_x) + (N_y - 2 n_y)} \Theta(n - n_x - n_y - n_z).\notag
\end{align}
Thus, if $\ket{\psi}$ is a permutationally invariant state, instead of computing the $4^N$ elements of the Pauli spectrum,  that is the set
\begin{equation}\label{eq:pauli-spectrum}
    \mathrm{Spec}\ket{\psi} = {\lbrace \lvert \braket{\psi | \hat P | \psi} \rvert \rbrace}_{\hat P \in \mathcal{P}_N}
\end{equation}
of $4^N$ real numbers $r_i \in [0, 1]$ given by the expectations of the Pauli strings on $\ket{\psi}$, one can just store the $\mathcal{D}$ expectation values of the symmetric representatives $\hat P(\vec{N})$ and their degeneracy $g(\vec{N})$ for any quadruple $\vec{N}$. Then, the SRE can be computed using Eqs.~\eqref{eq:sre} and~\eqref{eq:pi-pauli-sum}. In the Supplementary Material~\cite{Note3}, we discuss how permutational invariance simplifies the calculation of other measures of nonstabilizerness, such as the entanglement spectrum flatness~\cite{tirrito2023} and the stabilizer nullity~\cite{beverland2020}. This exponential simplification, arising from permutationally-invariant state tomography~\cite{moroder2012,novo2013}, is what allows us to exactly study the nonstabilizerness of systems of hundreds of qubits.

\textit{Lipkin-Meshkov-Glick model} --- Equipped with these tools, we study the nonstabilizerness in the spectrum of the LMG model~\cite{lipkin1965}. First, we focus on the ground state. We consider the Hamiltonian (in units of the exchange coupling $J$)
\begin{equation}\label{eq:lmg}
    \hat H / J = -(1-\xi) \, \hat S_z^{(N)} - (4\xi/N) {[\hat S_x^{(N)}]}^2,
\end{equation}
where $\xi \in [0, 1]$. The dimensionless transverse field, controlling the properties of the model, is $\gamma(\xi) \equiv (1-\xi) / 4\xi \in [0, \infty]$. This model features a second-order quantum phase transition at the critical value $\xi_c = 1/5$ ($\gamma_c = 1)$: for $\gamma < \gamma_c$ the ground state (GS) is ferromagnetic (aligned along $x$), while for $\gamma > \gamma_c$ the state is paramagnetic. The order parameter describing this symmetry-breaking transition is the magnetization in the $x$-direction $m_x$.
The half-system entanglement entropy of the ground state, $S_{N/2}$, displays a $\ln N$ divergence at the critical point of the mean-field (MF) transition, approaches the value $S_{N/2} = \ln 2$ in the zero-field limit, where the ground state is a GHZ state in the $x$ direction, and becomes zero in the opposite limit, where the GS is the product state polarized along $z$~\cite{latorre2005}. 

While the ground-state properties of the SRE can be inferred by mean-field theory~\cite{Note3}, states at finite energy density go beyond this description. 
The LMG Hamiltonian is made up of two terms: the first one has a $\mathrm{U}(1)$ symmetry structure, whereas the second one has an $\mathrm{SO}(2)$ symmetry. Together, they form a Hamiltonian with a $\mathrm{U}(2)$ algebraic structure. 
When $\xi > \xi_c$ this model is known to have an excited state quantum phase transition at finite energy density~\cite{santos2016}: for any value of $\xi_c \le \xi \le 1$ there exists a finite energy $E_\text{ESQPT}(\xi)$ around which the density of states is strongly peaked, i.\,e., the excited energy levels cluster around the separatrix line $E_\text{ESQPT}(\xi)$. In the following we will shift the Hamiltonian~\eqref{eq:lmg} by its ground-state energy $\hat H(\xi) \to \hat H(\xi) - E_\text{GS}(\xi)$. Then, the analytical expression of $E_\text{ESQPT}(\xi)$ can be derived in the large-$N$ limit with a semiclassical analysis and reads $E_\text{ESQPT}(\xi)/N = {(1-5\xi)}^2 / 16\xi$ for $\xi \ge \xi_c$. 
Some physical quantities, such as the longitudinal and transverse magnetization and the participation ratio in the $x$ or $z$ Dicke bases~\cite{santos2016},  are singular at the ESQPT and can thus be used as order parameters of the transition.
Interestingly, the study of the ESQPT yields a detailed fingerprint of the properties of the full many-body spectrum that goes beyond the mean-field description of the transition itself. 


We will explore the properties of the nonstabilizerness of the whole energy spectrum at finite energy density in a numerically exact fashion. We will show that its study reveals a very rich phenomenology, not grasped by the analysis of entanglement.

\begin{figure}[t]
    \centering
    \includegraphics[width = \columnwidth]{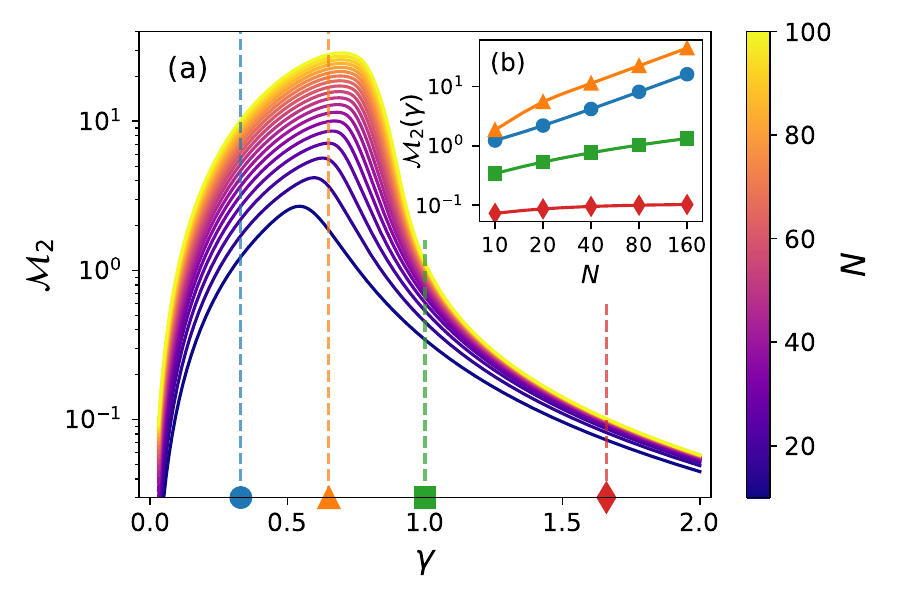}
    \caption{(a) Stabilizer R\'enyi entropy $\mathcal{M}_2$ of the ground state of the LMG model as a function of the transverse field strength $\gamma$, for several sizes (log-linear scale). (b) Scaling with $N$ of the SRE for the values of transverse field marked in Fig.~\ref{fig:sre-lmg}(a) (log-log scale).}
    \label{fig:sre-lmg}
\end{figure}

\textit{Ground state} --- In Fig.~\ref{fig:sre-lmg}(a), we report $\mathcal{M}_2(\ket{\text{GS}(\gamma)})$ as a function of the transverse field strength for system sizes ranging from $N = 10$ to $N = 100$ with increments of five qubits between subsequent curves. We see that the ground-state SRE is zero both for $\gamma = 0$ and for large $\gamma$ since both the GHZ state and the fully polarized state along the $z$ axis are stabilizer states, while it develops a peak for $0 < \gamma < 1$ that grows with $N$. The analysis of the SRE in the thermodynamic limit tells us that the peak position tends to $\gamma^* = 1/\sqrt{2}$, thus it does not approach the MF critical point, as opposed to the peak of entanglement~\cite{latorre2005,barthel2006,vidal2007,orus2008}.

\begin{table}[t]
\centering
\begin{tabular}{ccc}
    \toprule
    $\gamma$ & $\beta$ & $\sigma_\beta$ \\
    \midrule
    \hspace{3mm}0.33\hspace{3mm} & \hspace{3mm}0.973\hspace{3mm} & \hspace{3mm}0.002\hspace{3mm} \\
    \hspace{3mm}0.65\hspace{3mm} & \hspace{3mm}0.988\hspace{3mm} & \hspace{3mm}0.001\hspace{3mm} \\
    \hspace{3mm}1.00\hspace{3mm} & \hspace{3mm}0.401\hspace{3mm} & \hspace{3mm}0.006\hspace{3mm} \\
    \hspace{3mm}1.66\hspace{3mm} & \hspace{3mm}0.054\hspace{3mm} & \hspace{3mm}0.003\hspace{3mm} \\
    \bottomrule
\end{tabular}
\caption{Exponents of the scaling law $\mathcal{M}_2 \sim N^\beta$ and corresponding standard deviations, LMG model.}
\label{tab:lmg-exponents}
\end{table}

We can clearly distinguish two different scalings of the SRE with $N$, depending on the strength of the transverse field. For $\gamma < \gamma_c$, $\mathcal{M}_2$ is extensive and the stabilizer density $m_2 = \mathcal{M}_2 / N$ is constant. Adopting the same terminology used to describe entanglement phases, we identify this as a volume-law phase of magic. This result is surprising because it tells us that the ground state of the LMG model in the ferromagnetic phase cannot be prepared efficiently using Clifford gates, despite the fact that the large-$N$ limit of this model is exactly reproduced by mean-field theory. Instead, when $\gamma > \gamma_c$ we observe that the SRE becomes independent of $N$, signaling the onset of what can be called an ``area-law'' phase of magic. Here, the density of magic $m_2$ goes to zero with $N$, suggesting that large-$N$ ground states of the LMG model in this phase are essentially stabilizer states~\cite{tarabunga2023}. 


As shown in Fig.~\ref{fig:sre-lmg}(b) for systems up to $N = 160$ spins, the SRE follows a power-law scaling of the form $\mathcal{M}_2 \sim N^\beta$ with $\beta \ge 0$. The scaling exponent vanishes in the area-law phase, see red line with diamonds, and is close to one in the volume-law phase (see blue line with dots, orange line with triangles), see also Tab.~\ref{tab:lmg-exponents}. At the transition (green line with squares), the scaling with $N$ only looks sub-extensive due to finite-size effects, but mean-field theory~\cite{Note3} tells us that the SRE is extensive. Near the critical point, due to finite-size effects, we also observe a super-extensive region with an exponent $\beta > 1$, inconsistent with the mean-field predictions. The bound $\mathcal{M}_2 < N \ln 2$ is always satisfied~\cite{leone2022}.

\begin{figure}[t]
    \centering
    \includegraphics[width = 0.5\textwidth]{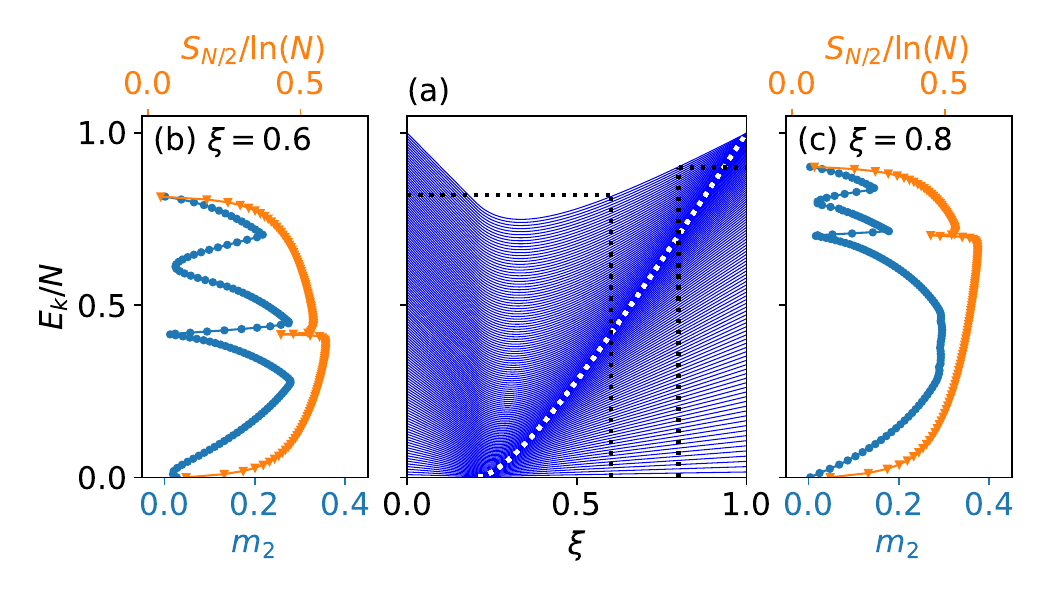}
    \caption{(a) Spectrum of the Hamiltonian~\eqref{eq:lmg}, for $N = 256$. (b) Entanglement entropy and density of magic for $\xi = 0.6$. (c) Same as (b), for $\xi = 0.8$.}
    \label{fig:lmg-spectrum-esqpt}
\end{figure}

\textit{Excited states} ---
We show the spectrum of the Hamiltonian~\eqref{eq:lmg} in Fig.~\ref{fig:lmg-spectrum-esqpt}(a) for $N = 256$. The separatrix line is marked with white dots and is clearly distinguishable by the peak in density of eigenstates. It has been shown that states exactly at the critical line are strongly localized around the ground state of the $\mathrm{U}(1)$ part of the Hamiltonian~\cite{santos2016}, i.\,e., the state  with all spins pointing up. This is a stabilizer state with zero magic. Close to the critical line states are generally entangled and have finite magic. Therefore, it is reasonable to expect that both the entanglement entropy and the SRE would sharply decrease in correspondence of the ESQPT, and indeed this is what we observe in Figs.~\ref{fig:lmg-spectrum-esqpt}(b,\,c), where we show the entanglement entropy and the density of magic $m_2 = \mathcal{M}_2 / N$ for all the eigenstates along the cuts $\xi = 0.6 $ and $\xi = 0.8$, respectively. Both these quantities sharply decrease when the cut intercepts the separatrix line, signaling the transition. Nevertheless, comparing the entanglement entropy and the SRE, we see that the latter has a richer structure, with cusps, local minima and local maxima as a function of the energy density. 

\begin{figure}[t]
    \centering
    \includegraphics[width = \columnwidth]{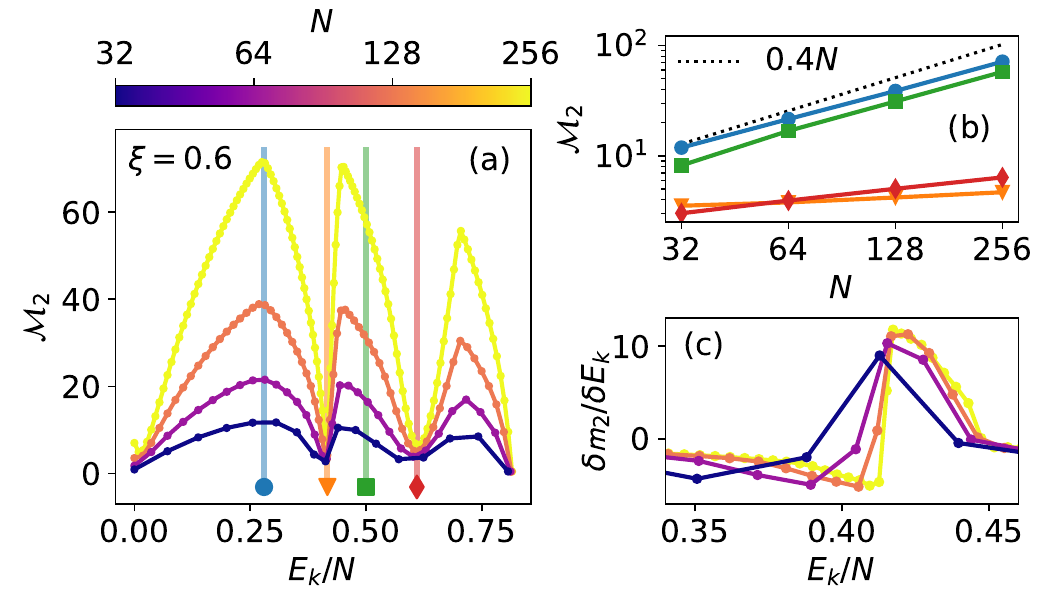}
    \caption{(a) SRE as a function of the energy density, along the cut $\xi = 0.6$, for several system sizes. (b) Scaling of $\mathcal{M}_2$ with $N$ for the energy densities marked in panel~{(a)} with the same symbols and colors. (c) Numerical derivative of the SRE around the critical density, for the same sizes as panel~{(a)}.}
    \label{fig:sre-esqpt-scaling}
\end{figure}

We focus on $\xi = 0.6$,  any $\xi > \xi_c$ shows the same features. In Fig.~\ref{fig:sre-esqpt-scaling}(a), we show the SRE along this cut for different sizes. There are regions where $\mathcal{M}_2$ is extensive and other regions where it is not. We recall that in permutationally symmetric systems the maximum amount of bipartite entanglement entropy is $S_\text{max} = \log D = \log (N + 1)$, whereas the maximum amount of magic is not restricted by symmetry: the maximum SRE is associated with a flat Pauli spectrum (except for the projection on the identity operator), which is allowed by permutation symmetry. In Fig.~\ref{fig:sre-esqpt-scaling}(b), we plot the scaling of $\mathcal{M}_2$ with $N$ for some energy densities, marked in panel~{(a)} using the same symbols. We see indeed regions where the exponent of the scaling law $\mathcal{M}_2 \sim N^\alpha$ is close to $\alpha \sim 1$, and other regions where $\alpha$ becomes close to zero. This happens, in particular, around the critical energy density, where, additionally, the SRE has a cusp. In the thermodynamic limit, its first derivative has a finite jump at $E_\text{ESQPT}$ as displayed in Fig.~\ref{fig:sre-esqpt-scaling}(c), where we plot the numerical derivative of the density of magic around the critical energy density and observe the development of a jump at the transition for large $N$.

\begin{figure}[t]
    \centering
    \includegraphics[width = \columnwidth]{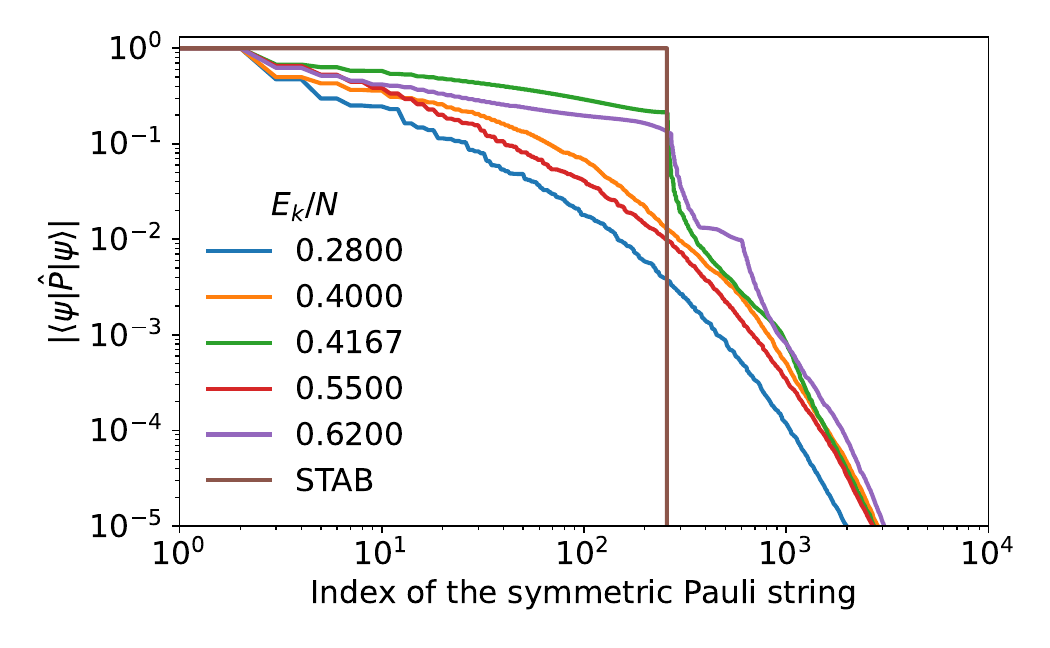}
    \caption{Pauli spectrum of different eigenstates of the LMG model at finite energy densities ($N = 256$).}
    \label{fig:pauli-spectrum}
\end{figure}

In order to better understand the differences between excited states with low or high magic, it is interesting to analyze the Pauli spectrum, Eq.~\eqref{eq:pauli-spectrum}, for different energy densities. For permutationally invariant systems, we only have to compute $\mathcal{D}$ distinct expectation values with their degeneracies $g(\vec{N})$. In Fig.~\ref{fig:pauli-spectrum}, we show the Pauli spectra of some eigenstates for $N = 256$ ($\mathcal{D} = 2862209$). In order to improve visualization, we sort the expectation values in decreasing order and only report one expectation value per symmetric representative without keeping track of its degeneracy. We observe that eigenstates in the volume-law phase of magic have a featureless Pauli spectrum. This has to be compared with the step-like spectra of states with minimal magic, which more closely resemble the spectrum of stabilizer states, which have nonzero expectation values ($\pm 1$) over exactly $2^N$ Pauli strings by definition. Notice, however, that in all cases the Pauli spectrum is significantly nonzero only for very few Pauli strings (compared to $\mathcal{D}$), which means one could easily combine the exponential reduction provided by permutation symmetry with the sampling techniques discussed in Refs.~\cite{lami2023,haug2023,tarabunga2023-2,lami2024} for an even more efficient estimation of the nonstabilizerness.

\textit{Conclusions} --- To summarize, exploiting the representation of the Pauli group in the Dicke basis, we studied the stabilizer R\'enyi entropy and other nonstabilizerness measures of permutationally symmetric systems, finding evidence of quantum magic phase transitions not detected in the entanglement.

\begin{acknowledgments}
{\em Acknowledgments} --- We would like to thank M.~Dalmonte, G.~Fux, A.~Hamma, G.~Lami, D.~Rattacaso, and E.~Tirrito for very helpful discussions and comments. G.\,P. acknowledges computational resources from the CINECA award under the ISCRA initiative, and from MUR, PON “Ricerca e Innovazione 2014-2020”, under Grant No.~PIR01\_00011 - (I.Bi.S.Co.). This work was supported by PNRR MUR project~PE0000023 - NQSTI, by the European Union’s Horizon 2020 research and innovation programme under Grant Agreement No~101017733, by the MUR project~CN\_00000013-ICSC (P.\,L.),  and by the  QuantERA II Programme STAQS project that has received funding from the European Union’s Horizon 2020 research and innovation programme under Grant Agreement No~101017733 (P.\,L.). This work is co-funded by the European Union (ERC, RAVE,~101053159) (R.\,F.). Views and opinions expressed are however those of the authors only and do not necessarily reflect those of the European Union or the European Research Council. Neither the European  Union nor the granting authority can be held responsible for them.
\end{acknowledgments}

%


\clearpage
\appendix
\begin{widetext}
\begin{center}
\textbf{\large \centering Supplementary Material: \\ Nonstabilizerness of Permutationally Invariant Systems}
\end{center}
\end{widetext}

\setcounter{equation}{0}
\setcounter{figure}{0}
\setcounter{table}{0}
\setcounter{page}{1}
\renewcommand{\theequation}{S\arabic{equation}}
\setcounter{figure}{0}
\renewcommand{\thefigure}{S\arabic{figure}}
\renewcommand{\thepage}{S\arabic{page}}
\renewcommand{\thesection}{S\arabic{section}}
\renewcommand{\thetable}{S\arabic{table}}
\makeatletter

\renewcommand{\thesection}{\arabic{section}}
\renewcommand{\thesubsection}{\thesection.\arabic{subsection}}
\renewcommand{\thesubsubsection}{\thesubsection.\arabic{subsubsection}}

\section{Symmetric representatives}

In this section, we expand the discussion of the main text by showing how to construct the matrix elements of the symmetric representatives of the Pauli group in the Dicke basis, given a quadruple $\vec{N} = \lbrace N_x, N_y, N_z, N_0 \rbrace $. We start by recalling the following property of Pauli matrices, which is a trivial consequence of Euler's formula:
\begin{equation}\label{eq:euler}
    \hat \sigma_j^{\alpha_j} = (-i) \exp\!\left[i \,(\pi/2) \, \hat \sigma_j^{\alpha_j}\right].
\end{equation}
Given a (non-symmetric) Pauli string $\hat P(\alpha_1, \dots, \alpha_N) = \bigotimes_{j = 1}^N \hat \sigma_j^{\alpha_j}$, we can rewrite it in the following alternative way, exploiting Eq.~\eqref{eq:euler} and the fact that Pauli matrices acting on different particles commute with each other:
\begin{align}
    &\hat P(\alpha_1, \dots, \alpha_N) = {(-i)}^{N - N_0} \exp\Bigl(i \,\frac{\pi}{2} \, \sum_{j \colon \alpha_j = x} \hat \sigma_j^{x}\Bigr) \notag \\& \qquad \times \exp\Bigl(i \,\frac{\pi}{2} \, \sum_{j \colon \alpha_j = y} \hat \sigma_j^{y}\Bigr) \exp\Bigl(i \,\frac{\pi}{2} \, \sum_{j \colon \alpha_j = z} \hat \sigma_j^{z}\Bigr).
\end{align}
We recognize, at the exponents, the components of the total spin operators built considering the particles on which the Pauli string acts with $X$, $Y$, or $Z$, respectively. As mentioned in the main text, the fundamental insight is that, for a permutationally invariant system, the specific lattice indices must not matter but only the number of times each gate appears in the Pauli string. This means that we can consider the total spin components of fully symmetric subsystems of $N_x$, $N_y$ and $N_z$ particles and write the symmetric representative $\hat P(\vec{N})$ as
\begin{equation}\label{eq:pauli-symmetric}
    \hat P(\vec{N}) = {(-i)}^{N-N_0} e^{i \pi \hat S_x^{(N_x)}} e^{i \pi \hat S_y^{(N_y)}} e^{i \pi \hat S_z^{(N_z)}}.
\end{equation}

In order to apply Eq.~\eqref{eq:pauli-symmetric} to a Dicke state, first we have to decompose the state to highlight its symmetric components. To do so, we can iteratively exploit the decomposition~\cite{latorre2005}
\begin{equation}\label{eq:dicke-decomposition-1}
    \ket{N, m} = \!\!\!\!\!\!\sum_{n_x = 0}^{\min(N_x, n)} \!\!\!\!\!\! \sqrt{p_{n, n_x}} \, \ket{N_x, n_x} \otimes \ket{N - N_x, n - n_x},
\end{equation}
with $p_{n,n_x} = \binom{N_x}{n_x} \binom{N - N_x}{n - n_x} / \binom{N}{n}$  which, after three applications, leads to the representation
\begin{align}\label{eq:dicke-decomposition}
    &\ket{N, m} = \sum_{n_x = 0}^{N_x} \sum_{n_y = 0}^{N_y} \sum_{n_z = 0}^{N_z} \sqrt{\frac{\binom{N_x}{n_x} \binom{N_y}{n_y} \binom{N_z}{n_z} \binom{N_0}{n - n_x - n_y - n_z}}{\binom{N}{n}}}\notag \\[1ex]
    & \quad \times \ket{N_x, n_x}\otimes\ket{N_y, n_y}\otimes\ket{N_z, n_z}\\[1ex]
    &\qquad\otimes\ket{N_0, n - n_x - n_y - n_z} \Theta(n - n_x - n_y - n_z),\notag
\end{align}
where $\Theta(x)$ is the Heaviside step function [$\Theta(x < 0) = 0$, $\Theta(x\ge0) = 1$]. At this point, we can use the following identities of the ($N_\alpha + 1$)-dimensional irreducible representations of the rotation matrices~\cite{sakurai} ($\alpha \in \lbrace x, y, z \rbrace$), which allow us to expand the three exponentials in Eq.~\eqref{eq:pauli-symmetric} in the bases of their respective spaces: 
\begin{subequations}
    \begin{gather}
        e^{i \pi \hat S_x^{(N_x)}} = {(i)}^{N_x} \sum_{n''_x = 0}^{N_x} \ket{N_x, n''_x}\bra{N_x, N_x - n''_x};\\[-1ex]
        e^{i \pi \hat S_y^{(N_y)}} = \sum_{n''_y = 0}^{N_y} {(-1)}^{n''_y} \ket{N_y, n''_y}\bra{N_y, N_y - n''_y};\\[-1ex]
        e^{i \pi \hat S_z^{(N_z)}} = {(i)}^{N_z} \sum_{n''_z = 0}^{N_z} {(-1)}^{n''_z} \ket{N_z, n''_z} \bra{N_z, n''_z}.
    \end{gather}
\end{subequations}
Putting everything together, we finally arrive at the matrix representation of $\hat P(\vec{N})$:
\begin{align}\label{eq:mel-pauli-supp}
    &\braket{N, m | \hat P(\vec{N}) | N, n} = {(-i)}^{N_y} \\[1ex] &\quad \times \sum_{n_x, n_y, n_z} {(-1)}^{n_y + n_z} \frac{\binom{N_x}{n_x} \binom{N_y}{n_y} \binom{N_z}{n_z} \binom{N_0}{n - n_x - n_y - n_z}}{\sqrt{\binom{N}{n}\binom{N}{m}}}\notag \\[1ex]
    &\qquad\times\delta_{m, n + (N_x - 2 n_x) + (N_y - 2 n_y)} \Theta(n - n_x - n_y - n_z).\notag
\end{align}
Each of the $\mathcal{D} = O(N^3)$ symmetric Pauli strings is a sparse matrix with at most $O(N^2)$ nonzero matrix elements. According to Eq.~\eqref{eq:mel-pauli-supp}, $O(N^2)$ operations are required to build each matrix element, since one of the three sums is cancelled by the Kronecker $\delta$. Therefore, with $ O(N^7)$ easily parallelizable operations, it is possible to compute the full representation of the Pauli group in the Dicke basis and store it into files for later access. At that point, the representation can be reused: in order to read it from memory, only $ O(N^5)$ operations are required since one no longer has to compute the double sum in Eq.~\eqref{eq:mel-pauli-supp}.

\section{Ferromagnetic p-spin model}

The LMG model can be generalized to systems with infinite-range $p$-body interactions with $p > 2$, giving rise to the ferromagnetic $p$-spin model~\cite{derrida1981,gross1984}. Its Hamiltonian reads
\begin{equation}\label{eq:pspin}
    \hat H / J = -\gamma \, \hat S_z^{(N)} - \frac{1}{2} {\left(\frac{2}{N}\right)}^{p-1} {[\hat S_x^{(N)}]}^p,
\end{equation}
from which one can re-obtain the LMG model by setting $p = 2$. As opposed to the LMG, the $p$-spin model with $p>2$ undergoes a first-order quantum phase transition in its ground state as a function of the transverse field $\gamma$, separating the paramagnetic from the ferromagnetic phase. The value of the critical field depends on $p$ and has been derived using mean-field theory for all values of $p$~\cite{bapst2012}. The bipartite entanglement entropy obeys an area law everywhere and is discontinuous at the transition~\cite{filippone2011}.

\begin{figure}[t]
    \centering
    \includegraphics[width = \columnwidth]{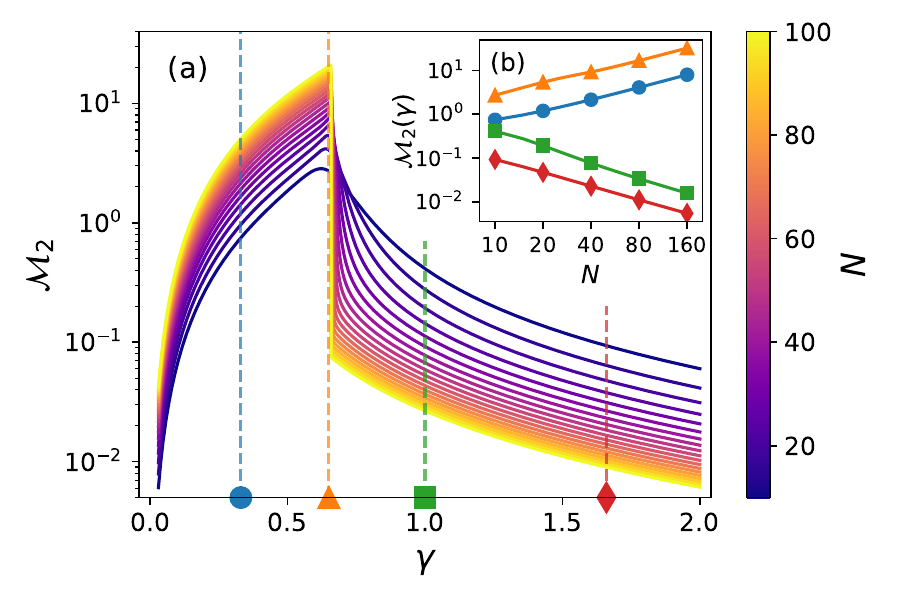}
    \caption{(a) Stabilizer R\'enyi entropy $\mathcal{M}_2$ of the ground state of the $p$-spin model ($p = 3$) as a function of the transverse field strength $\gamma$, for several sizes (log-linear scale). (b) Scaling with $N$ of the SRE for the values of transverse field marked in the main panel (log-log scale).}
    \label{fig:sre-pspin}
\end{figure}

We show the SRE of the ground state of the $p$-spin model with $p = 3$ in Fig.~\ref{fig:sre-pspin}(a) as a function of $\gamma$. First, we see that, in the ferromagnetic phase, $\mathcal{M}_2$ grows with $\gamma$ and $N$, until the critical point of the MF transition is reached. At that point, the SRE is discontinuous and suddenly jumps to a value close to zero. In the ferromagnetic phase, the SRE grows with $N$, scaling as $\mathcal{M}_2 \sim N^\beta$ with $\beta \approx 1$, i.\,e., it is extensive. By contrast, in the paramagnetic region, the SRE decreases with $N$: the scaling exponent $\beta$ becomes negative, as shown in Fig.~\ref{fig:sre-pspin}(b), for systems up to $N = 160$ spins, and in Tab.~\ref{tab:pspin-exponents}. The negative exponent tells us that the next-order term in powers of $1/N$ around the mean-field nonstabilizerness density decreases as $1/N^2$, as opposed to the LMG model shown in the main text, where the next-order term scales as $1/N$, see later.

\begin{table}[t]
\centering
\begin{tabular}{ccc}
    \toprule
    $\gamma$ & $\beta$ & $\sigma_\beta$ \\
    \midrule
    \hspace{3mm}0.33\hspace{3mm} & \hspace{3mm}0.943\hspace{3mm} & \hspace{3mm}0.003\hspace{3mm} \\
    \hspace{3mm}0.65\hspace{3mm} & \hspace{3mm}0.906\hspace{3mm} & \hspace{3mm}0.008\hspace{3mm} \\
    \hspace{3mm}1.00\hspace{3mm} & \hspace{3mm}-1.138\hspace{3mm} & \hspace{3mm}0.008\hspace{3mm} \\
    \hspace{3mm}1.66\hspace{3mm} & \hspace{3mm}-1.028\hspace{3mm} & \hspace{3mm}0.001\hspace{3mm} \\
    \bottomrule
\end{tabular}
\caption{Exponents of the scaling law $\mathcal{M}_2 \sim N^\beta$ and corresponding standard deviations, $p$-spin model.}
\label{tab:pspin-exponents}
\end{table}

\begin{figure}[tb]
    \centering
    \includegraphics[width = \columnwidth]{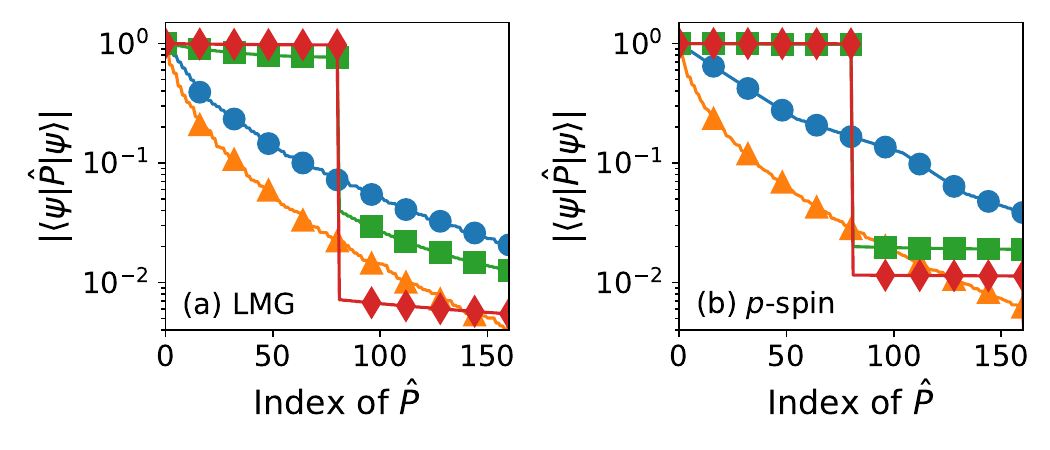}
    \caption{Pauli spectrum for $N = 80$ ($\mathcal{D} = 91881$), for (a) the LMG model and (b) the $p$-spin model ($p = 3$). Symbols refer to the values of $\gamma$ marked in Figs.~\ref{fig:sre-lmg} and~\ref{fig:sre-pspin}.}
    \label{fig:spectrum}
\end{figure}

In Fig.~\ref{fig:spectrum}, we show the Pauli spectra of the GS of the LMG and of the $p$-spin model for the values of the transverse field marked in Figs.~\ref{fig:sre-lmg} and~\ref{fig:sre-pspin}, for $N = 80$ ($\mathcal{D} = 91881$). In order to improve visualization, we sort the expectation values in decreasing order and only report one expectation value per symmetric representative without keeping track of its degeneracy. When $\gamma > \gamma_c$, we see that the ground state becomes close to a stabilizer state (which would be represented by a $1$-$0$ step function). Instead, when $\gamma \lesssim \gamma_c $, we see that the Pauli distribution develops a tail, which is significantly nonzero only for very few Pauli strings (compared to $\mathcal{D}$).

\section{Nullity and average entanglement-spectrum flatness}

In this section we show that other nonstabilizerness measures are simplified by permutational invariance.
The stabilizer nullity, 
\begin{equation}
    \nu(\ket{\psi}) = N - \log_2(\lvert \mathrm{Stab}\ket{\psi}\rvert),
\end{equation}
is another measure of nonstabilizerness that has the important property of being a provable nonstabilizerness monotone after measurements of Pauli operators~\cite{beverland2020}. Here, $\mathrm{Stab}\ket{\psi}$ denotes the subset of Pauli strings $\hat P$ such that $\hat P\ket{\psi} = \ket{\psi}$. Stabilizer states are such that $\lvert\mathrm{Stab}\ket{\psi}\rvert = 2^N$, hence their nullity is zero by definition. States with finite nonstabilizerness have $\lvert\mathrm{Stab}\ket{\psi}\rvert = 2^M$ for some $M < N$ and a nonzero nullity. States that are stabilized only by the identity have maximum nullity $\nu = N$. The stabilizer nullity is related to the stabilizer $k$-R\'enyi entropies $\mathcal{M}_k $ by the limit 
\begin{equation}
    \nu = \lim_{k\to\infty} (k - 1) \mathcal{M}_k
\end{equation}
and can also be computed from the Pauli spectrum, see Eq.~\eqref{eq:pauli-spectrum} of the main text. The number of ones in the Pauli spectrum is equal to $\lvert\mathrm{Stab}\ket{\psi}\rvert$~\cite{beverland2020}. Given the representation of the symmetric Pauli group, the stabilizer nullity can be computed as
\begin{equation}
    \nu(\ket{\psi}) = N - \log_2 \!\! \left[{\sum_{\vec{N}}^{\lvert \braket{\psi |\hat P(\vec{N})|\psi} \rvert = 1}} \!\!\!\!\!g(\vec{N})\right],
\end{equation}
where the sum is restricted to those $\vec{N}$ such that $\lvert \braket{\psi |\hat P(\vec{N})|\psi} \rvert = 1$, with $\mathrm{poly}(N)$ operations. The stabilizer nullity of the ground states of the analyzed models is always nonzero, except for $\gamma = 0$ (and $\gamma \to \infty$).

The average entanglement-spectrum flatness along orbits of the Clifford group $\mathcal{C}_N$~\cite{tirrito2023} is another recently proposed measure of nonstabilizerness. Compared to more intricate measures based on minimization of cost functions involving all stabilizer states, where symmetries are also equally beneficial~\cite{heinrich2019}, the entanglement-spectrum flatness shares with the SRE the fact that it is more easily computable, though, in general, with exponential effort. Moreover, it establishes a direct connection between entanglement response and nonstabilizerness. 

Given the pure-state density matrix $\hat \rho = \ket{\psi}\bra{\psi}$, we can consider a bipartition, $A + B$, of the $N$-qubit system made up of two subsets with $N_A$ and $N_B$ qubits, respectively, with $N_A + N_B = N$. The reduced density matrix of subsystem $A$ reads $\hat \rho_A = \mathrm{Tr}_B(\hat \rho)$, where $\mathrm{Tr}_B$ is the partial trace over $B$. The entanglement-spectrum flatness is defined as $\mathcal{F}_A(\ket{\psi}) = \mathrm{Tr}_A(\hat{\rho}_A^3) -  \mathrm{Tr}_A^2(\hat{\rho}_A^2)$. 

Nonstabilizerness is related to the average spectrum flatness along the Clifford orbit $\hat \Gamma \ket{\psi}$ (for all $\hat \Gamma \in \mathcal{C}_N$), denoted ${\langle \mathcal{F}_A(\hat \Gamma \ket{\psi})\rangle}_{\mathcal{C}_N}$. It has been noted that states with finite nonstabilizerness have a non-flat average entanglement spectrum, and viceversa. In Ref.~\cite{tirrito2023}, it was analytically proven that ${\langle \mathcal{F}_A(\Gamma \ket{\psi})\rangle}_{\mathcal{C}_N}  = c(D, D_A) \mathcal{M}_\text{lin}(\ket{\psi})$, where $\mathcal{M}_\text{lin} = 1 - D {\lVert \Pi(\ket{\psi}) \rVert}_2^2$ is the linear entropy, $\Pi(\ket{\psi})$ is the vector of probabilities of the Pauli strings, ${\lVert \cdot \rVert}_2$ is the Euclidean norm, and $c(D, D_A) $ is a coefficient that depends on the size of the partitions. The linear entropy is related to the 2-R\'enyi entropy by the relation $\mathcal{M}_2 = -\ln(1-\mathcal{M}_\text{lin})$, hence the average entanglement-spectrum flatness is directly related to the SRE, though it might be exponentially difficult to resolve due to the scaling of $c(D, D_A)$ with $N$.

\begin{figure}[tb]
    \centering
    \includegraphics[width = \columnwidth]{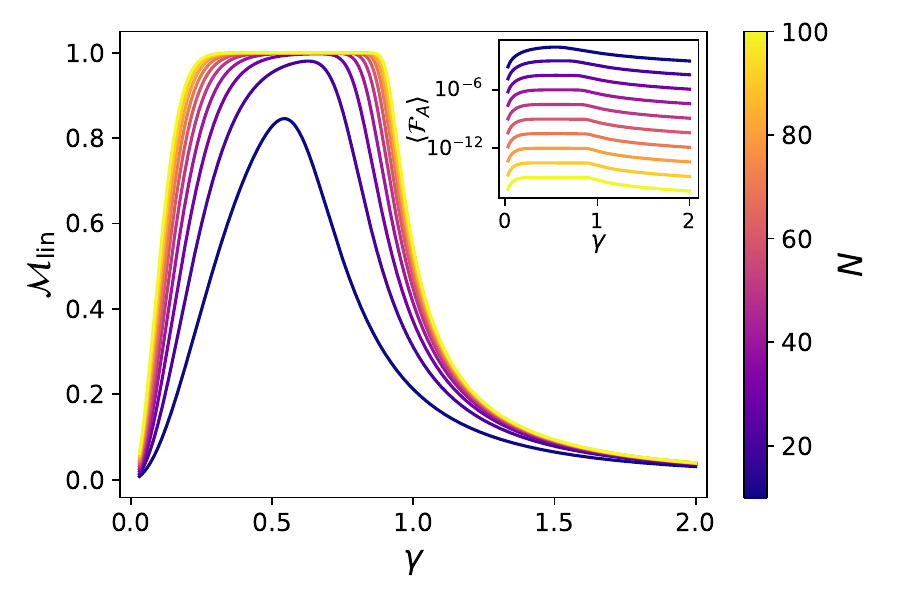}
    \caption{Average entanglement-spectrum flatness of the ground state of the LMG model as a function of the transverse field strength.}
    \label{fig:flatness}
\end{figure}
%
Therefore  the average entanglement-spectrum flatness can be computed as
\begin{equation}\label{eq:average-flatness}
    \langle \mathcal{F}_A \rangle = c(D, D_A) \mathcal{M}_\text{lin} \equiv \frac{(D^2-D_A^2)(D_A^2-1)}{(D^2-1)(D_A+2)D_A^2} \mathcal{M}_\text{lin},
\end{equation}
where $D = 2^N$ and $D_A$ is the size of the partition $A$. For a balanced bipartition, $D_A = \sqrt{D}$ and $c(D, D_A) \sim D^{-1}$ for large $N$, which means detecting the average entanglement-spectrum flatness is exponentially hard in $N$. For ease of visualization, we can instead refer to $\mathcal{M}_\text{lin}$. We plot it in Fig.~\ref{fig:flatness} for the LMG model for the same parameters and sizes analyzed in the main text. We can see that the flatness correctly detects the volume law phase of magic, where $\mathcal{M}_\text{lin}$ is close to one, while it becomes small in the area-law region or when $\gamma = 0$ and the ground state is a stabilizer. As shown in the inset, however, $\langle \mathcal{F}_A \rangle$ is more difficult to obtain for large $N$: already for these sizes, it can go down below machine precision.

\begin{figure}[tb]
    \centering
    \includegraphics[width = \columnwidth]{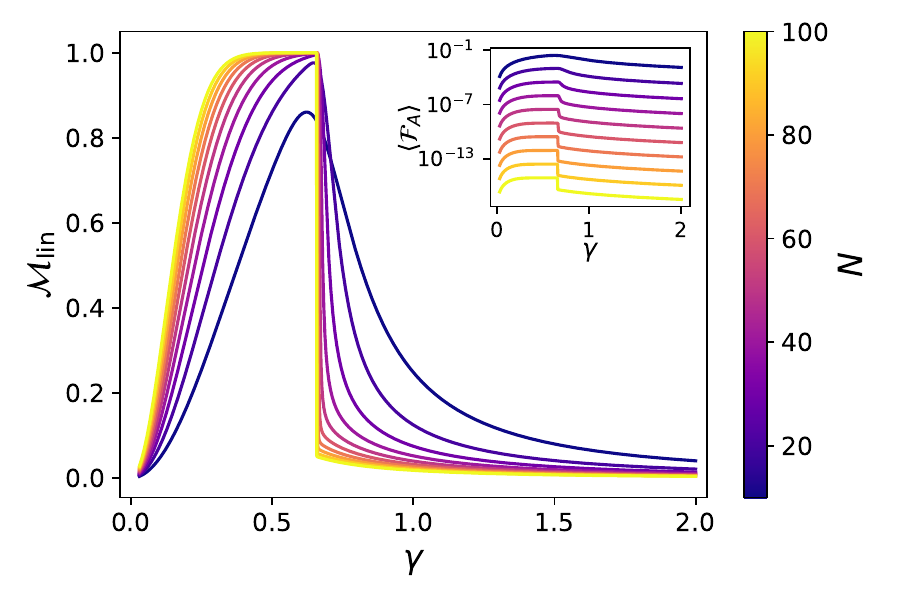}
    \caption{Average entanglement-spectrum flatness of the ground state of the $p$-spin model ($p = 3$) as a function of the transverse field strength.}
    \label{fig:flatness-pspin}
\end{figure}

Similar considerations apply to the $p$-spin model as well, see Fig.~\ref{fig:flatness-pspin}. Here, we also see that the average flatness becomes discontinuous at the transition for large $N$, but the issue of detecting an exponentially small quantity still remains.

\section{Mean-field stabilizer \texorpdfstring{R\'enyi}{Renyi} entropy}

For permutationally invariant systems, mean-field theory is exact in the thermodynamic limit. This allows deriving the analytical expression of the stabilizer R\'enyi entropy $\mathcal{M}_2$ of the ground states of the models considered for large $N$~\cite{tarabunga2023}. We report the general scheme in this section. 

The system's Hamiltonian is only expressed in terms of collective magnetization operators, $\hat H = \hat H(\lbrace \hat S_\alpha \rbrace)$. We can express the Hamiltonian in terms of the magnetization components $\hat m_\alpha = \hat S_\alpha / S$ with $S = N/2$ and define the Hamiltonian density $\hat h = \hat H / N$. Then, we notice that $[\hat m_\alpha, \hat m_\beta] \sim O(1/N)$, therefore, in the thermodynamic limit, the magnetization components behave like classical variables. In this limit, correlations between spins are negligible and the system's ground state is a tensor product of single-particle states aligned along the same angles $(\theta, \phi)$, a so-called spin-coherent state:
\begin{equation}\label{eq:spin-coherent}
    \ket{\theta,\phi} = \bigotimes_{k=1}^{N} \left(\cos \frac{\theta}{2} \ket{0}_k + \sin \frac{\theta}{2} e^{i \, \phi} \ket{1}_k\right).
\end{equation}
The expectation of $\hat h$ over the spin-coherent state gives the variational semiclassical energy $h(\theta, \phi)$, whose minimum corresponds to the ground-state energy in the thermodynamic limit and whose optimal angles $(\theta^*, \phi^*)$ allow writing the ground state as is Eq.~\eqref{eq:spin-coherent}. Then, since the ground state is a product state, we have that $\mathcal{M}_2 = N m_2$, with 
\begin{equation}\label{eq:sre-mf}
    m_2 = -\log \left[\frac{1 + \sin^4 \theta^* (\cos^4 \phi^* + \sin^4 \phi^*) + \cos^4 \theta^* }{2}\right].
\end{equation}

\begin{figure}[t]
    \centering
    \includegraphics[width=\columnwidth]{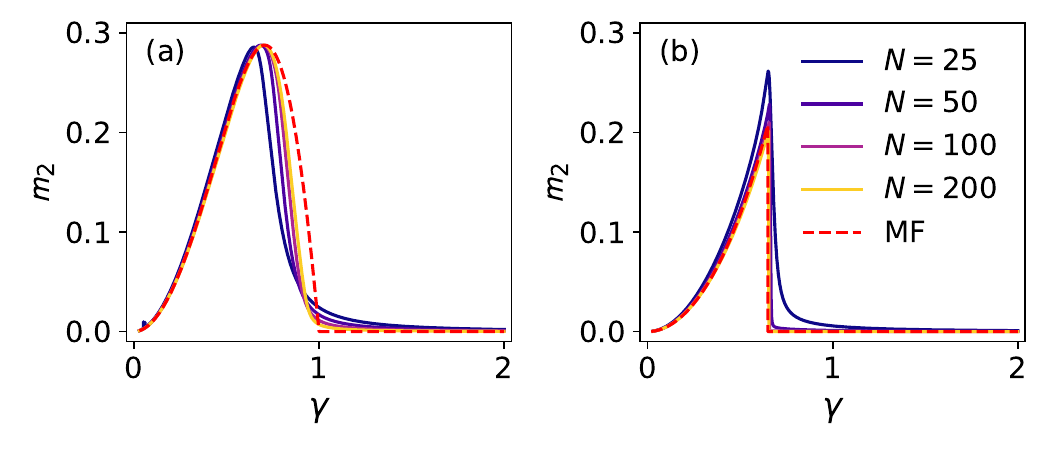}
    \caption{Comparison between the nonstabilizerness density $m_2$ evaluated with mean-field theory and with finite-size numerical simulations. (a) LMG model. (b) $p$-spin model.}
    \label{fig:sre-mf}
\end{figure}

For the LMG model, Eq.~\eqref{eq:lmg} of the main text, we observe that the Hamiltonian does not depend on $\hat S_y$, thus we get $\phi^* = 0 $ and write the semiclassical energy as 
\begin{equation}
    h(\theta) = -\frac{\gamma}{2} \cos \theta - \frac{1}{4} \sin^2 \theta.
\end{equation}
When $\gamma < \gamma_c = 1$, this function is minimized by $\theta^* = 0$; instead, when $\gamma > \gamma_c$, the minimum is $\theta^* = \arccos \gamma$. The order parameter $m_x = \sin\theta^*$ is continuous at the transition. The mean-field SRE of Eq.~\eqref{eq:sre-mf}, for $\gamma < \gamma_c$, reads
\begin{equation}
    m_2 = -\log\left(1+ \gamma^4-\gamma^2\right)
\end{equation}
and agrees with our numerical results, see Fig.~\ref{fig:sre-mf}(a). Its maximum value is found for $\gamma^* = 1/\sqrt{2}$.

For the $p$-spin model with $p = 3$, Eq.~\eqref{eq:pspin}, the semiclassical energy reads
\begin{equation}
    h(\theta) = -\frac{\gamma}{2} \cos \theta - \frac{1}{4} \sin^3 \theta
\end{equation}
and can be minimized as before. Following Ref.~\cite{bapst2012}, it can be shown that the order parameter is discontinuous at the transition and that the critical point is $\gamma_c = {(\sqrt{3}/2)}^3 \approx 0.6495$. The MF results agree with our large-$N$ simulations, see Fig.~\ref{fig:sre-mf}(b).

\begin{figure}[t]
    \centering
    \includegraphics[width = \columnwidth]{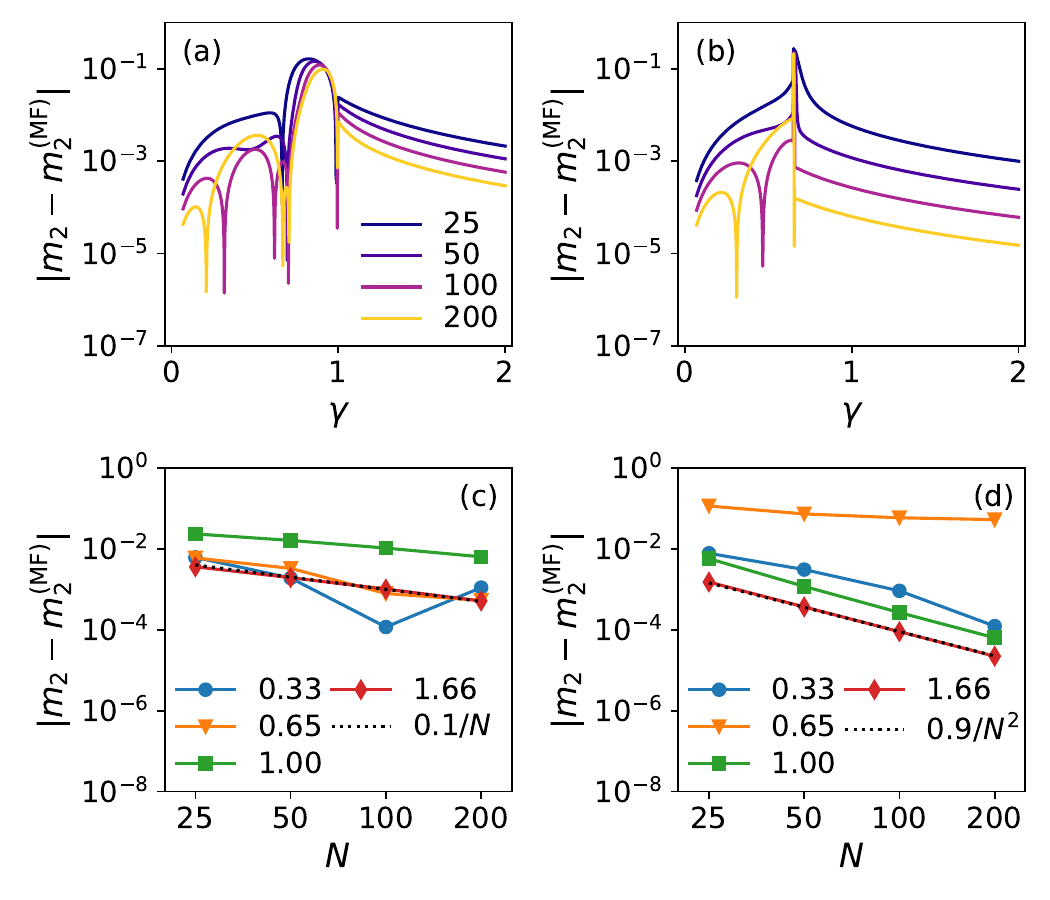}
    \caption{Next-to-leading order of the density of nonstabilizerness with respect to the mean-field solution, as a function of (a, b) $\gamma$ and (c, d) $N$, for the two analyzed models.}
    \label{fig:nextorder}
\end{figure}

It is also interesting to study the next-order corrections to the mean-field density of magic for the two analyzed models. In general,
\begin{equation}
    m_2(\gamma) = m_2^\text{(MF)}(\gamma) + \epsilon(\gamma, N).
\end{equation}
We numerically compute the difference $\epsilon$ and plot it in Figs.~\ref{fig:nextorder}(a,\,b), as a function of the transverse field for several system sizes. We can see that in the ferromagnetic region $\gamma < \gamma_c$ the curves generally do decrease with $N$, with some oscillations. In the paramagnetic region $\gamma > \gamma_c$, we see that the plotted quantity decreases with $N$ without oscillations. Moreover, we can clearly see that the scaling with $N$ is different in the two cases [the curves are more spread out in panel~{(b)} than in panel~{(a)}]. By focusing on some values of $\gamma$, we indeed observe different scalings with the system size. We plot $\epsilon $ as a function of $N$ in Figs.~\ref{fig:nextorder}(c,\,d). In the paramagnetic region, e.\,g., $\gamma = 1.66$, we find $\epsilon_\text{LMG} \approx 1/10N$ and $\epsilon_\text{$p$-spin} \approx 0.9 / N^2$. While we do not have an analytical explanation of these scalings, they allow us to understand the two very different scalings of the SRE with $N$ in the paramagnetic phase of the two analyzed models, Figs.~\ref{fig:sre-lmg} and~\ref{fig:sre-pspin}.

\end{document}